\documentclass[prl,aps, twocolumn,floatfix,longbibliography]{revtex4-1}
\usepackage[utf8]{inputenc}
\usepackage{natbib}
\usepackage{graphicx}
\usepackage{color}
\usepackage[tbtags]{amsmath}
\usepackage{amssymb}
\usepackage{gensymb}
\usepackage{float}

\newcommand{\la}{\langle}
\newcommand{\ra}{\rangle}

\newcommand{\lp}{\left(}
\newcommand{\rp}{\right)}

\renewcommand{\vec}[1]{{\bf #1}}

\renewcommand{\phi}{\varphi}
\renewcommand{\epsilon}{\varepsilon}

\begin{document}

\title{Plasmonic nonreciprocity driven by band hybridization in moir{\'e} materials}

\author{Micha{\l} Papaj}
\thanks{These authors contributed equally to this work.}
\author{Cyprian Lewandowski}
\thanks{These authors contributed equally to this work.}
\affiliation{Department of Physics, Massachusetts Institute of Technology, Cambridge, Massachusetts 02139, USA}

\begin{abstract}
We propose a new current-driven mechanism for achieving significant plasmon dispersion nonreciprocity in systems with narrow, strongly hybridized electron bands. The magnitude of the effect is controlled by the strength of electron-electron interactions $\alpha$, which leads to its particular prominence in moir{\'e} materials, characterized by  $\alpha \gg 1$.  Moreover, this phenomenon is most evident in the regime where Landau damping is quenched and plasmon lifetime is increased. The synergy of these two effects holds a great promise for novel optoelectronic applications of moir\'e materials.
\end{abstract}

\maketitle

\textit{Introduction}.---
Time-reversal symmetry breaking leads to the emergence of unidirectional modes in platforms such as the quantum Hall systems \cite{PhysRevLett.45.494, PhysRevB.25.2185, PhysRevB.29.1616}, the quantum anomalous Hall materials \cite{PhysRevLett.61.2015,Yu61,PhysRevLett.106.166802,PhysRevLett.113.147201,Chang167}, or the topological photonic crystals \cite{RevModPhys.91.015006,10.1038/nphoton.2014.248,10.1038/nphys3796,10.1038/s41566-017-0048-5,SUN201752}. However, such modes, while holding an exceptional promise for the development of new devices, often require very specific experimental conditions, such as strong magnetic fields, significant magnetic impurity doping or a large, macroscopic size of a device. Frequently, such systems cannot be easily coupled to electromagnetic radiation, limiting their experimental utility. Moreover, they are not easily susceptible to miniaturization necessary for the technological applications, which usually benefit from nanoscale on-chip integration. 

One of the alternative platforms in which nonreciprocity is highly sought-after are the 2D surface plasmons \cite{10.1038/nphoton.2015.201,Song201519086,PhysRevX.8.021020,PhysRevB.93.041413,PhysRevLett.117.196803,10.1038/ncomms13486,PhysRevLett.118.245301,PhysRevLett.100.023902,borgnia2015quasirelativistic,Duppen_2016,Bliokh:18,PhysRevB.92.195429,acsphotonics.8b00987}, collective charge density modes of fundamental importance in controlling light-matter interactions \cite{10.1038/nphoton.2012.262,10.1038/nphys2615,10.1038/nature01937}.
These quasiparticles can be excited using electromagnetic radiation and are an essential ingredient in developing optoelectronic devices. While nonreciprocity in the plasmon dispersion, $\omega_{p}({\vec{q}}) \neq \omega_{p}({-\vec{q}})$, can be induced using magnetic field \cite{PhysRevLett.54.1710,PhysRevLett.54.1706, HEITMANN1986332, PhysRevB.28.4875}, 2D plasmons also allow for an appealing alternative based on driving electric current through the devices - the so-called plasmonic Doppler effect \cite{PhysRevLett.71.2465,borgnia2015quasirelativistic, PhysRevB.92.195429,acsphotonics.8b00987,Bliokh:18,Duppen_2016}. The essence of this phenomenon boils down to a simple Galilean transformation that distinguishes plasmons moving along and against the electric current. Electron flow modifies the plasmon dispersion with a correction, $\Delta \omega_{p}^{(c)}{\sim}{\vec{u}\cdot\vec{q}}$, proportional to the drift velocity $\vec{u}$ and plasmon momentum $\vec{q}$. This current-induced nonreciprocity is the conventional plasmonic Doppler effect. 

\begin{figure*}[!tb]
    \centering
    \includegraphics[width=0.8\linewidth]{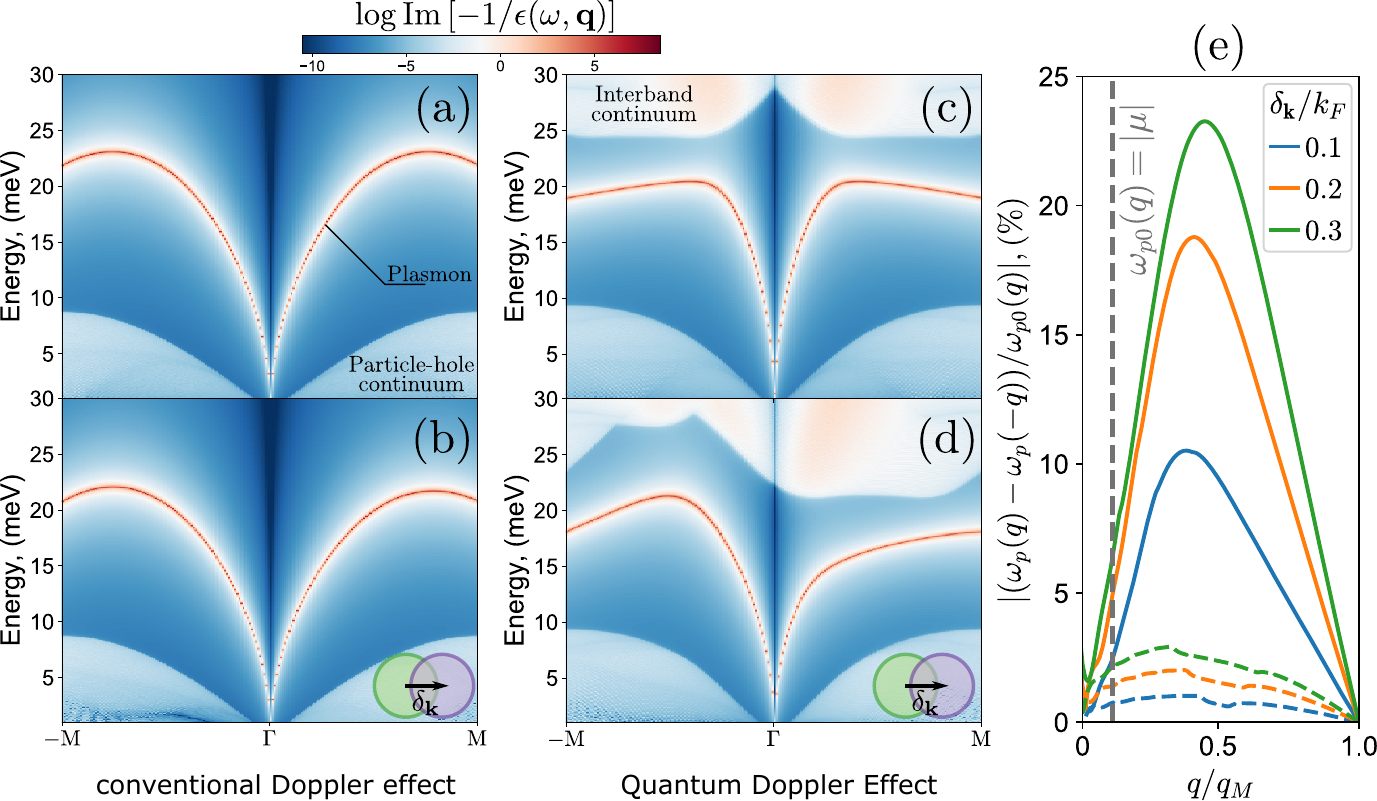}
    \caption{(a-d) Electron loss functions of the narrow band tight-binding models. (a, b) Unhybridized band model (a) without and (b) with electric current at $T=0$. The conventional Doppler effect imposes only a small change on the plasmon dispersion. (c, d) Hybridized band model (c) without and (d) with electric current at $T=0$. Quantum Doppler effect results in a strong plasmon dispersion asymmetry.
     (e) A comparison of relative dispersion asymmetry at several drift velocity values. In hybridized uystem (solid) strong asymmetry develops, which is purely a consequence of a non-vanishing interband wavefunction overlap, in contrast to an unhybridized system showing only a conventional Doppler effect (dashed). Here $q_M$ is the length of $\Gamma-M$ vector of the moir{\'e} Brillouin zone.}
    \label{fig:fig_1}
\end{figure*}

 Unfortunately, even in pristine graphene samples, the drift velocity is a small fraction of Fermi velocity $v_F$ \cite{doi:10.1063/1.3483130,nl204545q}. Therefore the relative magnitude of the Doppler effect \cite{borgnia2015quasirelativistic}
\begin{equation}
\label{eq:relative_doppler_shift_vanilla}
\frac{\Delta \omega_{\rm p}^{(c)}({\vec{q}})}{\omega^0_{p}({\vec{q}})}\sim\frac{1}{\alpha} \frac{u}{v_F} \frac{\omega^0_{p}({\vec{q}})}{|\mu|},\quad\omega^0_{p}({\vec{q}}) = \sqrt{4 \alpha |\mu| v_F q}
\end{equation}
is a small correction on the order of $\sim 3\%$ to the graphene plasmon dispersion in the absence of electron drift, $\omega^0_{p}({\vec{q}})$ \cite{PhysRevB.75.205418,Wunsch_2006,nl201771h, PhysRevB.80.245435,doi:10.1063/1.2891452}, as shown in Fig.~\ref{fig:fig_1}(a,b). Here $|\mu|$ is the Fermi energy and $\alpha = e^2/\hbar \kappa v_F$ characterizes the strength of the electronic interactions in a dielectric medium with a relative permittivity $\kappa$. Since in the most common scenarios $\alpha \sim 1$ (e.g. monolayer graphene), its presence in the drift-free part of the plasmon dispersion means that $\omega_{p}^0(\vec{q}) < |\mu|$, which is an additional limitation in the attempts to observe the conventional Doppler effect.

\begin{figure}[!tb]
    \centering
    \includegraphics[width=\linewidth]{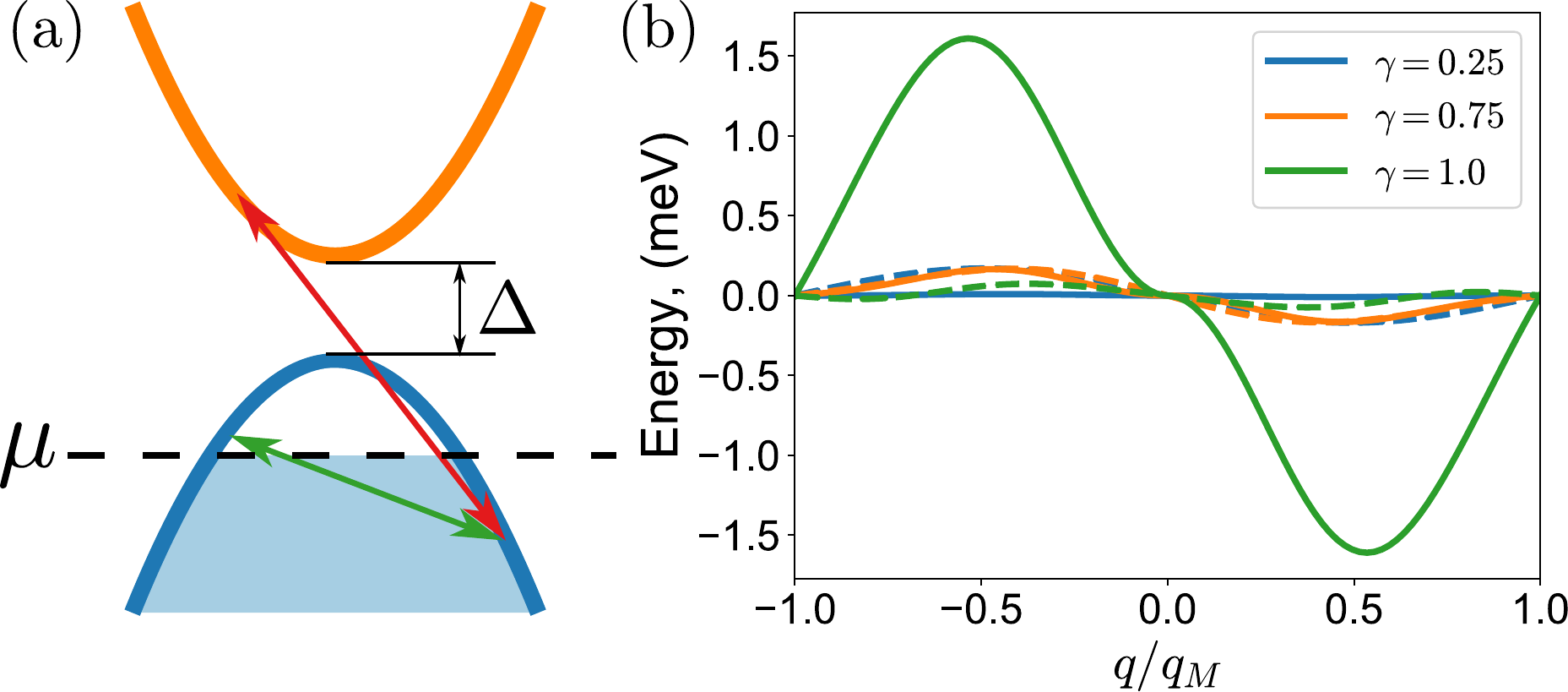}
    \caption{(a) The  polarization function contribution of the interband transitions (red), unlike that of the intraband transitions (green), is suppressed for $\omega$ smaller than the band gap $\Delta = 2\sqrt{\Delta_d^2+\Delta_h^2}$. (b) Shift of the plasmon dispersion obtained from the tight-binding model due to the quantum (solid) and the conventional (dashed) Doppler effect for different degree of band hybridization $\gamma$. Here $\Delta_h = \gamma \Delta$ and $\Delta_d = \sqrt{1-\gamma^2} \Delta$ to keep the gap $\Delta$ constant.}
    \label{fig:fig_2}
\end{figure}

Here we show that strongly hybridized, narrow band materials characterized by $\alpha \gg 1$,  can host a new, fundamentally quantum in nature, source of plasmonic nonreciprocity. In such case, the asymmetry of the plasmon dispersion (as demonstrated in Fig.\ref{fig:fig_1}(c,d)) is strongly enhanced by an additional factor of $\alpha$
\begin{equation}
\Delta \omega_{\rm p}^{(q)}({\vec{q}}){\sim} \alpha \frac{\Delta_h^2 v_F q}{|\mu|^3} \vec{u} \cdot \vec{q}\,, \, \frac{\Delta \omega_{\rm p}^{(q)}({\vec{q}})}{\omega^0_{p}({\vec{q}})}{\sim}\frac{u}{v_F} \frac{\Delta_h^2  \omega^0_{p}({\vec{q}})}{|\mu|^3} \frac{q}{k_F}\,, \label{eq:quantum_doppler_shift}
\end{equation}
where $\Delta_h$ is the strength of hybridization between the two bands that opens up a gap between them (see Fig.\ref{fig:fig_2}a). Here, the relative frequency shift is amplified by the effective fine structure factor $\alpha$ unlike in the conventional Doppler effect $\Delta \omega_{\rm p}^{(c)}(\vec{q})$.  
The origin of this new effect can be traced back to the hybridization effects in electronic bandstructure. When plasmon frequencies exceed the chemical potential, a regime guaranteed by the strong interactions $\alpha \gg 1$ \cite{Lewandowski20869}, the new mechanism dominates over the conventional one, leading to a strong enhancement of plasmonic nonreciprocity.

While this new source of bandstructure-driven nonreciprocity is a general consequence of band hybridization, the necessary ingredient ensuring a drastic increase in the effect's magnitude
is the presence of strong electron-electron interactions. A natural platform with these attributes are the moir{\'e} materials, such as the twisted bilayer graphene (TBG) \cite{cao1,cao2,Yankowitz1059} or the ABC stacked trilayer graphene (TLG) \cite{chenEvidenceGatetunableMott2019}. This enhancement of electron-electron interactions is due to the emergence of a superlattice with a period much larger than the atomic spacing of the original crystal. Such a large lattice constant results in a small Brillouin zone, giving rise to a set of extremely narrow minibands with bandwidths on the order of tens of meV \cite{Bistritzer12233,PhysRevB.86.125413}. Therefore, moir{\'e} materials are in many ways an ideal realization of a strongly correlated system: the same sample can display a record-low density superconductivity \cite{cao2,Yankowitz1059, chenSignaturesTunableSuperconductivity2019}, a correlated insulating state \cite{cao1, chenEvidenceGatetunableMott2019, luSuperconductorsOrbitalMagnets2019b}, or an interaction-driven ferromagnetism \cite{Sharpe605,Serlineaay5533}. These narrow bands also offer a key advantage to plasmonics: narrow-band plasmons can rise above particle-hole continuum and thus quenching Landau damping \cite{Lewandowski20869,khaliji2019plasmons}. These characteristics make moir{\'e} materials a perfect platform to realize nonreciprocal plasmons with long lifetimes.

In this work we focus specifically on TLG as it features a single separated flat band that can be tuned using external electric field \cite{chenSignaturesTunableSuperconductivity2019,chenEvidenceGatetunableMott2019}. We employ a continuum model \cite{chenEvidenceGatetunableMott2019,zhangBandStructureABC2010, koshinoTrigonalWarpingBerry2009} to perform a material-realistic calculation of the plasmon dispersion. These simulations show a significant plasmonic nonreciprocity, exceeding that predicted due to the conventional plasmon Doppler effect, and thus demonstrating moir{\'e} materials as a promising optoelectronics platform. 

\textit{Minimal bandstructure model}.---
To elucidate the microscopic origins of the new nonreciprocity mechanism, we develop a minimal model capturing the essential features of the complicated moir{\'e} bandstructures relevant to the plasmonic Doppler effect. We use a toy-model Hamiltonian $H = H_0 + H_d + H_h$, where:
\begin{equation}
H_0 = \frac{k^2}{2m} \, \sigma_{z}, \quad H_d = \Delta_d \sigma_z,\quad H_h = \Delta_h \sigma_x
\end{equation}
$H_0$ consists of two parabolic bands that can be thought of as coming from a tight-binding model. Here $m$ is the effective mass large enough such that the plasmons extend above the intraband particle-hole continuum of each band, and $\sigma_{x,y,z}$ are the Pauli matrices. To describe the energy gap separating the flat band from the rest of the moir{\'e} mini-bands, we use two mechanisms: $H_d$, a trivial displacement-field-like gap, and $H_h$, a band hybridization term. We label the electron energies and their Bloch eigenstates as $E_{s, \vec{k}}$ and $\psi_{s, \vec{k}}$ respectively, with a schematic bandstructure shown in Fig.\ref{fig:fig_2}a. We place the Fermi energy $\mu$ inside the valence band so that it qualitatively corresponds to the flat band of TLG.

\textit{Plasmons in narrow-band materials}.---
Collective charge modes correspond to the nodes of the dynamical dielectric function $\epsilon(\omega, \vec{q})=1-V_{\vec{q}}\Pi(\omega,\vec{q})$, where $V_{\vec{q}} = 2\pi e^2/\kappa q$ is the Coulomb potential. We calculate the electron polarization function $\Pi(\omega,\vec{q})$ within the random phase approximation \cite{mahan2000many-particle}
\begin{equation}
\Pi(\omega,\vec{q}) = 4 \sum_{\vec{k},s,s'} \frac{ (f_{s,\vec k+\vec q}-f_{s',\vec k})F^{s s'}_{\vec{k}+\vec{q},\vec{k}}}{E_{s,\vec{k}+\vec{q}}-E_{s',\vec{k}}-\omega-i0}
,
\label{eq:pol_tight_binding}
\end{equation}
where $\sum_{\vec{k}}$ denotes integration over the Brillouin zone and the indices $s,s'$ run over electron bands. The factor of $4$ accounts for the four-fold spin/valley degeneracy mimicking the degeneracy of the TLG superlattice. Here $f_{s,\vec k}$ is the Fermi-Dirac distribution, and $F^{ss'}_{\vec{k}+\vec{q},\vec{k}} = |\la \psi_{s,\vec{k}+\vec{q}} | \psi_{s', \vec{k}} \ra|^2$ describes the overlap between the Bloch eigenstates.

\textit{Origins of plasmonic nonreciprocity}.---
We now focus on explaining the behavior of the plasmon modes in this toy-model with an electron carrier drift present in the system. As the interband terms in the polarization function will be suppressed by the large denominator on the order of the bandgap energy scale, it is therefore sufficient to focus only on the intraband contribution. To that end we expand the intraband term in powers of $1/\omega$ to obtain
\begin{equation}
\Pi(\omega,\vec{q}) \approx \frac{A_1(\vec{q})}{\omega} + \frac{A_2(\vec{q})}{\omega^2} + \frac{A_3(\vec{q})}{\omega^3}+\dots\,.
\end{equation}
The coefficients in the above expansion are
\begin{equation}
A_n(\vec{q}) = 4\sum_\vec{k} \Tilde{f}_\mathbf{k} \lp F^{--}_{\vec{k},\vec{k}+\vec{q}} \Delta E_{\vec{k}+\vec{q},\vec{k}}^{n-1}  - F^{--}_{\vec{k},\vec{k}-\vec{q}} \Delta E_{\vec{k},\vec{k}-\vec{q}}^{n-1} \rp\label{eq:a_n_starting_point}
\end{equation}
with $\Delta E_{\vec{k},\vec{k'}}^{n} \equiv \lp E_{-,\vec{k}} - E_{-, \vec{k'}} \rp^{n}$ corresponding to the $n$th power of the energy difference of the intraband transitions, and $\Tilde{f}_\mathbf{k} \equiv f_{-,\vec k - m\vec{u}}$ denoting the drift-modified distribution function as described in Supplemental Materials\footnote{See Supplemental Material where additional details of analytical derivation and numerical calculation are discussed.}. These expressions rely on the Fermi energy $\mu$ placement in the valence band $s=-$ and hence the conduction band being completely unoccupied at low temperatures.

Now we analyze the most insightful regime of $|\mu| \gg \Delta_d, \Delta_h$. We expand the band overlap factors and the energy differences in the small-$q$ limit and then focus only on the leading $\vec{k}$ behavior of $A_n(\vec{q})$. We begin with the $A_1(\vec{q})$ coefficient, obtaining
\begin{align}
   A_1(\vec{q}) &\approx -\frac{2}{\pi} \frac{\Delta_h^2 u q^3 \cos(\theta_u)}{|\mu|^3 }\,,  \label{eq:a1_final}
\end{align}
where we approximated Fermi energy as $|\mu| \approx k_F^2/2m$ and $\theta_u$ is the angle between $\vec{q}$ and $\vec{u}$.  As expected, in the absence of drift current, $\vec{u}=0$, the time-reversal symmetry is preserved and the odd $1/\omega$ powers in expansion of $\Pi(\omega,\vec{q})$ vanish \cite{lan84}. Furthermore, if there is no hybridization between the bands, $\Delta_h=0$, the $1/\omega$ contribution to the polarization clearly vanishes.
 
Following the same approach, we now evaluate $A_2(\vec{q})$ and $A_3(\vec{q})$. To the leading order in $\vec{q}$ we can set the band overlap factors in Eq.\eqref{eq:a_n_starting_point} as unity, finding:
\begin{equation}
A_2(\vec{q}) \approx \frac{2}{\pi} |\mu| q^2,  \quad A_3(\vec{q}) \approx -\frac{4}{\pi} u \cos\theta_u |\mu|  q^3\,. \label{eq:a2_a3_final}
\end{equation}
The $q$ dependence of the $A_n(\vec{q})$ coefficients is easily understood. This is because the lowest possible contribution to the polarization function is always of the order ${\sim}q^2$\cite{mahan2000many-particle} and thus the first term which can be an odd function of the angle $\cos(\theta_u)$ has to scale as $q^3$.

We are now in position to obtain the plasmon dispersion $\omega_\text{p}(\vec{q})$ using the cubic equation
\begin{align}
0&=\omega^3-\frac{2\pi \alpha v_F}{q}\left( A_1(\vec{q}) \omega^2+A_2(\vec{q})\omega+A_3(\vec{q})\right)\,\label{eq:cubic}
\end{align}
with $v_F = k_F/m$. Solving this equation perturbatively in the powers of the electron drift velocity $u$ we find the plasmon dispersion as
\begin{align}
\omega_{\text{p}}(\vec{q}) &\approx \sqrt{4\alpha |\mu| v_F q} -2 \alpha \frac{\Delta_h^2 v_F q }{|\mu|^3}\vec{u}\cdot\vec{q} -  \vec{u} \cdot \vec{q}\,,\label{eq:master_equation}
\end{align}
which is the central result of our work. It is the last two terms in the above expression that are behind the plasmonic nonreciprocity in the presence of electron drift.

\begin{figure*}
    \centering
    \includegraphics[width=0.8\linewidth]{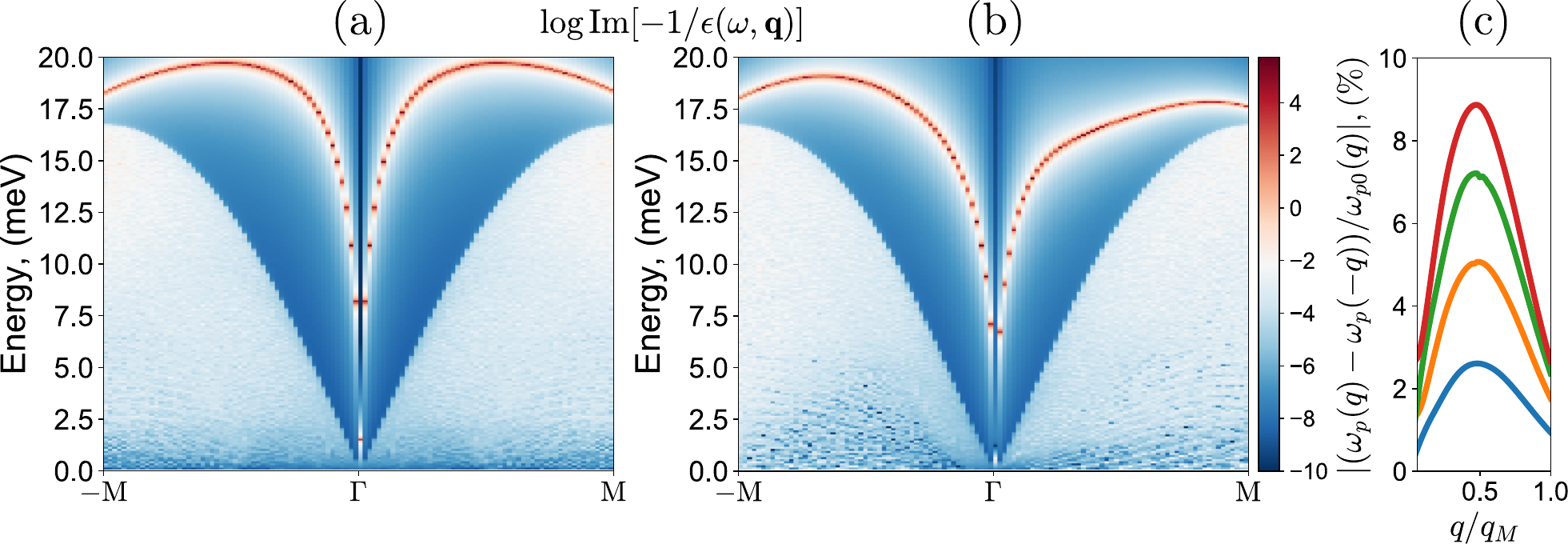}
    \caption{ Electron loss function in TLG (a) without and (b) with an applied electric current. The plasmon dispersion exhibits strong nonreciprocity under the Fermi surface shift $\delta_\vec{k} = 0.2 q_M$. (c) Relative nonreciprocity for several values of $\delta_\vec{k} = c\, q_M$, with $c=0.05, 0.1, 0.15, 0.2$ and $q_M = 0.24 ~\rm{nm}^{-1}$.}
    \label{fig:TLG_results}
\end{figure*}

\textit{Quantum Doppler Effect}.---
The second term in the Eq. \eqref{eq:master_equation} is the new source of plasmonic nonreciprocity, which dominates in narrow-band materials. To see this we analyze the system parameters' dependence of the Doppler corrections. 

The conventional Doppler shift (last term) depends only on the drift velocity $u$ and thus its magnitude is only weakly tunable. For $q{\sim}{k_F}$ it is a fraction of the chemical potential, 
\begin{equation}
\Delta \omega_{p}^{(c)}\approx u k_F \approx {\frac{u}{v_F} |\mu|}\,, \quad |\mu|\approx v_F k_F\,,
\end{equation}
as the drift velocity $u$ is always smaller than the Fermi velocity $v_F$.
This contrasts the quantum contribution, the second term in Eq.~\eqref{eq:master_equation}, where the effect's magnitude can be drastically increased by effective fine structure constant $\alpha$. At momenta $q{\sim}{k_F}$ it is
\begin{equation}
\Delta \omega_{p}^{(q)}\approx 2 \alpha \frac{u}{v_F} \frac{\Delta_h^2 v_F^2 k_F^2 }{|\mu|^3} \approx {2 \alpha \frac{u}{v_F} \frac{\Delta_h^2}{|\mu|}}\,,
\end{equation}
and thus a large $\alpha$ offers a parametric increase of the effect. This is exactly the behavior we expect in narrow-electron band systems where $\alpha \gg 1$.

To further demonstrate this point we perform numerical calculations based on the narrow-band tight-binding model described in Supplementary Materials\cite{Note1}. The plasmonic dispersion in the absence of electric current is shown in Fig.\ref{fig:fig_1}(a, c) for two parameter regimes: one with strongly hybridized bands due to $\Delta_h$ term, and the other with decoupled bands simply displaced by a finite energy $\Delta_d$. In both scenarios the parameters are chosen to keep the same bandwidth and bandgap of $10$ meV. Both cases exhibit qualitatively similar behavior - plasmons' dispersions settle between the intra- and inter-band particle-hole continua as guaranteed by $\alpha \gg 1$ \cite{Lewandowski20869}. However, when electric current is introduced, the striking difference between them is immediately apparent. While in the unhybridized case the observed nonreciprocity is minute, Fig.\ref{fig:fig_1}(b), in the system with hybridized bands a strong asymmetry in plasmon dispersion arises, see Fig.\ref{fig:fig_1}(d). The nonreciprocity can be quantified by the dispersion asymmetry between the $\vec{q}$ and $-\vec{q}$ modes
$(\omega_p(\mathbf{q}) - \omega_p(-\mathbf{q}))/\omega_p^0(\mathbf{q})$, displayed in Fig.\ref{fig:fig_1}(e) for both cases.  While the conventional effect is present in both cases, the calculation in a strongly hybridized system reveals a remarkable, order of magnitude enhancement over the unhybridized one in agreement with the analytical calculation. This comparison between the conventional and the quantum Doppler effect is further exemplified by the crossover from the strongly to weakly hybridized system shown in Fig.\ref{fig:fig_2}(b).  We vary the degree of band hybridization $\gamma$ while keeping constant the bandwidths and bandgaps, and plot $\Delta \omega_{p}^{(c)} = A_3/2A_2$ and $\Delta \omega_{p}^{(q)}= \pi \alpha v_F A_1/q$, with the latter clearly dominating when the bands are strongly hybridized.

We highlight that the $A_1(\vec{q})$ term responsible for the quantum Doppler effect is not just a special feature of our model, but rather is universal to any system with hybridized bands. In fact, the $1/\omega$ term appears also in the graphene Doppler shift calculations \cite{borgnia2015quasirelativistic,PhysRevB.92.195429}, but because the relevant plasmon frequencies are smaller or comparable to the Fermi energy $\omega_p \lesssim |\mu|$, the $A_1(\vec{q})$ term is suppressed by a small ratio of $q^2/k_F^2$. More generally, the origins of the $A_1(\vec{q})$ coefficient stem from a finite difference of the band overlap functions. This overlap measures the extent to which wave functions' spectral weight at different momenta come from the same bands. It is strongly dependent on the band hybridization and reaches unity far from the band crossing points as $|\mu|$ becomes larger. Indeed it is the relation between the chemical potential and the plasmon frequency which determines the crossover to the regime in which quantum contribution dominates, 
\begin{equation}
\label{eq:greater_doppler}
\Delta \omega_{\rm p}^{(q)}(\vec{q}) \gtrsim  \Delta \omega_{\rm p}^{(c)}(\vec{q}) \Rightarrow \omega_p^0(\vec{q}) \gtrsim |\mu|,
\end{equation}
as indicated in the Fig.\ref{fig:fig_1}(e). 

\textit{Doppler effect in moir{\'e} materials}.---
We turn now to a particular material realization of this phenomenon - ABC stacked trilayer graphene. To obtain electron bands and Bloch wavefunctions we perform a realistic material calculation using the continuum model introduced in Ref.\cite{chenSignaturesTunableSuperconductivity2019, chenEvidenceGatetunableMott2019,koshinoTrigonalWarpingBerry2009,zhangBandStructureABC2010}\cite{Note1}. With that model we numerically evaluate the dielectric function and determine the resulting plasmon dispersion.  

Fig.~\ref{fig:TLG_results}(a, b) demonstrate the plasmon dispersion in TLG without and with electric current, respectively. As in the tight binding model, an asymmetry $\omega_{\rm p}(\vec{q})\neq \omega_{\rm p}(-\vec{q})$ develops due to the flowing electric current. In analyzing this figure it is insightful to compare it with the Fig.\ref{fig:fig_1}(c, d) which reproduces qualitative features of the TLG calculation. Most crucially, we see a plasmon mode which rises above the particle-hole continuum and once the mode $\omega_{\rm p}(\vec{q})$ exceeds the Fermi energy $|\mu|$ a strong nonreciprocity in its dispersion develops. This behavior is to be expected on the basis of the analysis leading to Eq.\eqref{eq:greater_doppler}.

In Fig.~\ref{fig:TLG_results}(c) we see by evaluating the nonreciprocity measure that even for a realistic bandstructure and drift velocities the induced nonreciprocity is a significant correction to the plasmon dispersion exceeding conventional Doppler effect predictions. We underline again that the enhancement of the Doppler effect from Eq.\eqref{eq:quantum_doppler_shift} is a general feature of systems with narrow, strongly hybridized bands and thus not limited to TLG - we expect it to be present, and perhaps be even more pronounced, in other materials with these characteristics.

\textit{Summary and outlook}.---
A key feature shared by many moir{\'e} materials is their remarkable flat electron bands with extremely low Fermi velocity  and, therefore, exceptionally large effective fine structure constant  $\alpha$ values. In this work we showed how such strong interactions can lead to a new, significant source of plasmon nonreciprocity. Our results have immediate consequences of both practical and fundamental importance. First of all, they open a pathway to development of optoelectronic devices with suppressed backscattering \cite{RevModPhys.91.015006,PhysRevLett.102.213903, nphys1893,ncomms13657,10.1038/nphoton.2013.185}, for example plasmonic isolators based on Mach-Zehnder interferometers \cite{doi:10.1063/1.3127531,doi:10.1063/1.126284}, making them a valuable addition to the nanophotonics toolbox. Moreover, the drift-based mechanism enables a highly tunable electrical control of nonreciprocity on a nanoscale by simply controlling the current flow in the device. This on-chip compactness and tunability are in striking contrast to the mechanisms that employ the magnetic-based approaches. Finally, introducing a nonreciprocity to the dispersion of plasmons with quenched Landau damping is particularly appealing, as it paves a way towards a practical realization of various theoretical predictions, such as the Dyakonov-Shur instability \cite{PhysRevLett.71.2465}, that were previously limited by the plasmonic lifespan. As the collective modes in the moir{\'e} materials are actively searched for using near-field optical microscopy techniques \cite{10.1038/nature11254, 10.1038/nature11253, PhysRevLett.119.247402, hesp2019collective}, this work can open new prospects for both fundamental and practical applications of moir{\'e} plasmons.

\begin{acknowledgements}
We thank Leonid  Levitov for drawing our attention to the concept of plasmonic Doppler effect and Ali Fahimniya for useful discussions. M. P. was supported by DOE Office of Basic Energy Sciences under Award DE-SC0018945. C. L. acknowledges support from the MIT Physics graduate program and the STC Center for Integrated Quantum Materials, NSF Grant No. DMR-1231319.
\end{acknowledgements}

\bibliography{references}

\pagebreak
\widetext

\setcounter{equation}{0}
\renewcommand{\theequation}{S\arabic{equation}}
\renewcommand{\thefigure}{S\arabic{figure}}

\section*{Supplemental Material for ``Plasmonic nonreciprocity driven by band hybridization in moir{\'e} materials"}

\section{Additional details of the analytical derivation}
In this section of the Supplemental Materials we provide the full details of the analytical derivation of the plasmonic Doppler effect. We focus on a two band toy-model that reproduces the physics of narrow-band plasmons in moir{\'e} materials. In the main text we defined the toy-model Hamiltonian to be $H=H_0+H_d + H_h$ with:
\begin{equation}
H_0 = \frac{k^2}{2m} \, \sigma_{z}, \quad H_d = \Delta_d \sigma_z,\quad H_h = \Delta_h \sigma_x
\end{equation}
Its eigenvalues are:
\begin{equation}
E_{s, \vec{k}} = s \sqrt{\Delta_h^2 +  H_{\vec{k}}^2}, \quad H_{\vec{k}} \equiv k^2/2m + \Delta_d
\end{equation}
and the corresponding Bloch eigenstates: 
\begin{equation}
\psi_{s, \vec{k}} = \frac{1}{\sqrt{2 E_{s, \vec{k}}\lp E_{s, \vec{k}} + H_{\vec{k}}\rp}} \begin{pmatrix}H_{\vec{k}} + E_{s, \vec{k}} \\ \Delta_h\end{pmatrix}.
\end{equation}
with a momentum $\vec{k}$ and a band index $s=\pm$ corresponding to the conduction ($+$) and valence ($-$) bands. The overlap between these Bloch eigenstates is given by:
\begin{equation}
F^{ss'}_{\vec{k}+\vec{q},\vec{k}} = |\la \psi_{s,\vec{k}+\vec{q}} | \psi_{s', \vec{k}} \ra|^2 = \frac{\lp (E_{s,\vec{k}+\vec{q}}+H_{\vec{k}+\vec{q}}) (E_{s',\vec{k}}+H_{\vec{k}}) + \Delta_h^2 \rp^2}{4 E_{s,\vec{k}+\vec{q}} E_{s',\vec{k}} (E_{s,\vec{k}+\vec{q}}+H_{\vec{k}+\vec{q}}) (E_{s',\vec{k}}+H_{\vec{k}})} \,.
\end{equation}

We pause to clarify the electric and magnetic field nature of the plasmon modes that are found through the nodes of the dielectric function $\epsilon(\omega,\vec{q})$
\begin{equation}
    \epsilon(\omega,\vec{q}) = 1 - V_{\vec{q}}\Pi(\omega,\vec{q})\,,\label{eq:app_dielectric_fun_def}
\end{equation}
where $V_{\vec{q}}=2\pi e^2/q$ is the 2D Coulomb interaction and $\Pi(\omega,\vec{q})$ is the dynamical polarization function. As discussed in the main text we approximate the polarization function by its RPA expression
\begin{equation}
\Pi(\omega,\vec{q}) = 4 \sum_{\vec{k},s,s'} \frac{ (f_{s,\vec k+\vec q}-f_{s',\vec k})F^{s s'}_{\vec{k}+\vec{q},\vec{k}}}{E_{s,\vec{k}+\vec{q}}-E_{s',\vec{k}}-\omega-i0}\,.
\label{eq:app_pol_tight_binding}
\end{equation}

The modes obtained from the nodes of $\epsilon(\omega,\vec{q})$ correspond precisely to the frequency of the longitudinal charge oscillations in a solid: plasmons. However, what is experimentally relevant are not these longitudinal charge oscillations, but rather closely related oscillations of the transverse magnetic and electric field components. More precisely, they are the transverse magnetic surface plasmon polariton (TM-SPP) waves with a slightly more involved equation defining their dispersion\cite{PhysRevB.80.245435,PhysRevB.92.195429}.  However under the assumption that the retardation effects can be neglected, $v_F \ll c$, the defining relation for dispersion of TM-SPP waves reduces to seeking the nodes of the dielectric function, Eq.\eqref{eq:app_dielectric_fun_def}. In the Section \ref{app:full_wave} of the Supplemental Materials we demonstrate the validity of the non-retardation assumption as well as we present the behavior of the electric and magnetic fields that comprise these surface plasmons.

We return to the analysis of the polarization function. To that end we rewrite the polarization function from Eq.\eqref{eq:app_pol_tight_binding} by carrying out a replacement $\vec{k}+\vec{q} \to \vec{k}$ and $s\to s'$ in the first fraction with the $f_{s,\vec k+\vec{q}}$ distribution function. This yields an expression
\begin{equation}
\label{eq:app_pi_starting_point}
\Pi(\omega,\vec{q}) = 4 \sum_{\vec{k},s,s'} f_{s,\vec k - m \vec{u}}\left(\frac{ F^{s s'}_{\vec{k},\vec{k}-\vec{q}}}{E_{s,\vec{k}}-E_{s',\vec{k}-\vec{q}}-\omega-i0}-\frac{ F^{s s'}_{\vec{k},\vec{k}+\vec{q}}}{E_{s',\vec{k}+\vec{q}}-E_{s,\vec{k}}-\omega-i0}\right)\,,
\end{equation}
which we proceed to expand in the long wavelength limit.  In the small-$q$ limit the energy associated with the intraband transitions will be always smaller than the frequency $\omega$, while the energy of the interband transitions will be always larger than $\omega$. As discussed in the main text, it is therefore sufficient to focus on the intraband contribution to the polarization function only as the interband terms will be suppressed by the large denominator on the order of the bandgap energy scale.

In the above Eq.\eqref{eq:app_pol_tight_binding} to account for the flow of the electric current in the system we modified the Fermi-Dirac distribution \cite{borgnia2015quasirelativistic,PhysRevB.92.195429}. To the leading order in the strength of the electric field $\vec{E}$, it induces a shift of the Fermi sea by a momentum $\delta_\vec{k} = -e \vec{E} \tau$. Here $\tau$ is a characteristic momentum relaxation timescale which underlying microscopic form may be highly nontrivial. We sidestep this difficulty by parametrizing the momentum shift instead as $\delta_\vec{k} = -m \vec{u}$ with $\vec{u}$ being the experimentally determined drift velocity. The effect of the electron drift onto the polarization function is then simply given by a replacement $f_{s,\vec k} \to f_{s,\vec k + \delta_\vec{k}}$, made in both distribution functions in Eq.\eqref{eq:app_pol_tight_binding} above.

To obtain a closed form of the coefficients $A_n(\vec{q})$ introduced in the main text
\begin{equation}
A_n(\vec{q}) = 4\sum_\vec{k} \Tilde{f}_\mathbf{k} \lp F^{--}_{\vec{k},\vec{k}+\vec{q}} \Delta E_{\vec{k}+\vec{q},\vec{k}}^{n-1}  - F^{--}_{\vec{k},\vec{k}-\vec{q}} \Delta E_{\vec{k},\vec{k}-\vec{q}}^{n-1} \rp\,,\label{eq:app_a_n_starting_point}
\end{equation}
it is necessary to understand the practical implications of the limit of $|\mu| \gg \Delta_d, \Delta_h$. The largest energy scale that controls the behavior of the expansion coefficients $A_n(\vec{q})$ is the Fermi energy. As such, we can therefore expand the band overlap factors and the energy differences in the small-$q$ limit and then subsequently focus only on the leading $\vec{k}$ behavior of the $A_n(\vec{q})$ coefficients. In practice this translates to simply approximating the exact electron energies $E_{s,\vec{k}}$  as parabollically dispersing carriers. This yields the following expression for the energy difference 
\begin{equation}
\Delta E_{\vec{k}+\vec{q},\vec{k}} \approx \frac{-q^2}{2m}- \frac{k q }{m} \cos\theta
\end{equation}
and the band overlap factors
\begin{align}
\label{eq:app_bcf_apparent}
F^{--}_{\vec{k}+\vec{q},\vec{k}} \approx 1-\frac{ 4\Delta_h^2 m^2 q^2 \cos\theta}{k^6} \left(\cos\theta+\frac{ q}{k}(1-4\cos^2\theta)\right)\,.
\end{align}
In the above we introduced an angle $\theta$ between the vectors $\vec{k}$ and $\vec{q}$. Note that in the limit of $\Delta_h\to 0$ intraband overlap approaches $F^{--}_{\vec{k},\vec{k}+\vec{q}} \to 1$. This is to be expected as in the limit of $\Delta_h\to 0$ the two bands of the toy-model $H$ become unhybridized - the matrix $H$ is diagonal. We pause here to note that the apparent divergence of the band overlap factor, Eq.\eqref{eq:app_bcf_apparent}, as momentum $k \to 0$ is a consequence of the assumption $|\mu| \gg \Delta_d, \Delta_h$. This is a justified approximation as in practice, upon summation over the BZ, the contribution of the band overlap factors to the polarization function will be dominated by the momenta $k$ close to the Fermi momentum $k_F$. 

We now demonstrate the explicit evaluation of the $A_n(\vec{q})$ coefficients by starting with the $A_1(\vec{q})$ term. The difference of the band overlap factors in Eq.\eqref{eq:app_a_n_starting_point},
\begin{align}
    &F^{--}_{\vec{k},\vec{k}+\vec{q}} - F^{--}_{\vec{k},\vec{k}-\vec{q}} \approx \frac{8 \Delta_h^2 m^2 \cos\theta\left(4\cos^2\theta-1\right)}{k^7} q^3\, \label{eq:bcf_difference} 
\end{align}
projects only odd $\vec{k}\cdot \vec{q}$ components. For the integration over the direction of $\vec{k}$ in Eq.\eqref{eq:app_a_n_starting_point} not to vanish the drift-modified Fermi-Dirac term $\Tilde{f}_\mathbf{k}$ has to similarly contribute an odd harmonic of $\vec{k}$. As required when $\vec{u} = 0$, that is TRS is not broken, the Fermi surface is an even harmonic of $\vec{k}$ and hence the $1/\omega$ term is absent \cite{lan84}. With $\vec{u} \neq {0}$ however we expect the Fermi-Dirac distribution to develop odd harmonics linear in $\vec{u}$ that are centered near the Fermi momentum $k_F$. At zero temperature we can model such shifted Fermi-Dirac distribution $\Tilde{f}_\mathbf{k}$ as an $\theta$ angle dependent Heaviside function
\begin{equation}
\Tilde{f}_\mathbf{k}=-\Theta\left( k_F + m u\cos(\theta_u-\theta) - k\right).
\end{equation}
We choose a coordinate system such that $\theta$, $\theta_u$ are the angles between $\vec{q}$ and vectors $\vec{k}$, and $\vec{q}$ and $\vec{u}$, respectively. Here the negative sign in front of the Heaviside function stems from setting the charge neutrality point at $\mu = 0$. This yields 
\begin{align}
   A_1(\vec{q}) &\approx -\frac{2}{\pi} \frac{\Delta_h^2 u q^3 \cos(\theta_u)}{|\mu|^3 }  \label{eq:app_a1_final}
\end{align}
for the $A_1(\vec{q})$ term where we approximated Fermi energy as $|\mu| \approx k_F^2/2m$. Following the same approach, the next two coefficients $A_2(\vec{q})$ and $A_3(\vec{q})$ can be obtained as well. Using the parabolic energy dispersion approximation and setting the band overlap factors in Eq.\eqref{eq:app_a_n_starting_point} as unity we find
\begin{equation}
A_2(\vec{q}) \approx \frac{2}{\pi} |\mu| q^2,  \quad A_3(\vec{q}) \approx -\frac{4}{\pi} u \cos\theta_u |\mu|  q^3\,. \label{eq:app_a2_a3_final}
\end{equation}

As discussed in the main text, with the $A_n(\vec{q})$ coefficients known in a closed form, we are now in position to obtain the plasmon dispersion $\omega_\text{p}$ analytically. To that end we seek zeros of the dielectric function which in terms of the $A_n(\vec{q})$ coefficients becomes now a cubic equation
\begin{align}
   0&=\omega^3-\frac{2\pi \alpha v_F}{q}\left( A_1(\vec{q}) \omega^2+A_2(\vec{q})\omega+A_3(\vec{q})\right)\,\label{eq:app_cubic}
\end{align}
with $v_F = k_F/m$. Since $A_1(\vec{q})$ and $A_3(\vec{q})$ are both functions of the drift velocity $u$, they are a parametrically small correction to the dispersion as compared to the $A_2(\vec{q})$ term. We note in passing that the $A_1(\vec{q})$ coefficient enters as a prefactor of a larger power of $\omega$ than the term $A_2(\vec{q})$ which defines the unperturbed plasmon energy scale $\omega^0_{\rm p}(\vec{q})$. This is in contrast to the $A_3(\vec{q})$ term responsible for the conventional Doppler effect, which enters as a lower power of $\omega$ and hence is suppressed by the large $\omega^0_{\rm p}(\vec{q})$ energy scale. Solving the equation Eq.\eqref{eq:app_cubic} perturbatively in the powers of the electron drift velocity $u$ we find the plasmon dispersion as
\begin{align}
\label{eq:app_plasmon_disperion_general}
\omega_{\text{p}}(\vec{q}) &\approx \sqrt{2 \pi \alpha v_F A_2 / q} + \pi \alpha v_F A_1 / q + \frac{A_3}{2 A_2} +\mathcal{O}(u^2)\,,
\end{align}
where $A_1(\vec{q})$ and $A_3(\vec{q})$ are both linear functions of the drift velocity. 
Using the $A_n(\vec{q})$ coefficients from Eq.\eqref{eq:app_a1_final} and Eq.\eqref{eq:app_a2_a3_final} in the above solution gives the expression
\begin{align}
\omega_{\text{p}}(\vec{q}) &\approx \sqrt{4\alpha |\mu| v_F q} -2 \alpha \frac{\Delta_h^2 v_F q }{|\mu|^3}\vec{u}\cdot\vec{q} -  \vec{u} \cdot \vec{q}\,,\label{eq:app_master_equation}
\end{align}
discussed in the main text.

We conclude this Section by drawing attention to the $q$ dependence of the analytic expressions for both Doppler effect contributions: $\omega_{p}^{(q)}$ and $\omega_{p}^{(c)}$, the last two terms of Eq.\eqref{eq:app_master_equation} respectively. The quantum Doppler effect enters at a higher power of momentum, $\omega_{p}^{(q)} \propto q^2$, making it at first glance seem to be smaller than the conventional Doppler contribution,  $\omega_{p}^{(c)} \propto q$. This difference in the powers of momentum $q$ in both terms can be traced back to the division by the $A_2(\vec{q})$ term, c.f. Eq.\eqref{eq:app_plasmon_disperion_general}, as both $A_1(\vec{q})$, given by Eq.\eqref{eq:app_a1_final}, and $A_3(\vec{q})$, given by Eq.\eqref{eq:app_a2_a3_final}, enter as the same power of momentum $q$. This lack of division in the $\omega_{p}^{(q)}$ term by the factor $A_2(\vec{q})$, which sets the scale for the main drift-free part of the plasmon dispersion $\omega_{p}^0$, is ultimately behind the parametric enhancement of the quantum Doppler effect contribution by the large effective fine structure constant $\alpha$. As demonstrated with tight-binding simulations discussed in the main text, this large factor of $\alpha$ overcomes the nominal suppression stemming from the additional factor of $q/k_F$ present in the quantum Doppler contribution term for plasmon frequencies larger than the chemical potential, $\omega_{p}(\vec{q}) \gtrsim|\mu|$. 

\section{The tight-binding model\label{app:tight_binding}}
To compare and contrast the quantum and conventional plasmonic Doppler effect we use a simple tight-binding model that retains some features of the true TLG bands. The key properties of the bandstructure that control the behavior of the collective modes are the electron band's bandwidth (a natural bandstructure cut-off), momentum scale, and similar symmetry properties. For this purpose we use a nearest neighbor tight-binding model on a triangular lattice with two orbitals per site. For clarity, we use hoppings of equal magnitude, but opposite sign ($t_1 = -t_2 = t$), arriving at:
\begin{align}
H_{TB} &= - 2 t \left(\cos(k_x a) + 2 \cos\left(\frac{k_x a}{2}\right) \cos\left(\frac{k_y a \sqrt{3}}{2}\right) - 3 \right) \sigma_z \notag \\
&+\Delta_d \sigma_z + \Delta_h \sigma_x
\label{eq:app_H_tight_binding}
\end{align}
Here the lattice constant $a = 15 \mathrm{nm}$ and $\sigma_i$ are the Pauli matrices. In Fig.1(a), (b) the hopping magnitude is $t = 2.3\, \mathrm{meV}$, while the gap parameters are $\Delta_d = -3\, \mathrm{meV}$ and $\Delta_h = 10\, \mathrm{meV}$, and the calculation is performed for $\mu=-14.5\, \mathrm{meV}$ and $\kappa=4.5$. For the comparison of the magnitude of nonreciprocity between hybridized and unhybridized cases we also use $t = 1.1\, \mathrm{meV}$, $\Delta_d = 10\, \mathrm{meV}$ and $\Delta_h = 0\, \mathrm{meV}$ to obtain the unhybridized bands of comparable bandwidth. In Fig.2(b), where we show the quantum and conventional contributions to the plasmonic nonreciprocity we keep the bandwidth $W$ and bandgap $\Delta$ at 10 meV while we tune between two types of the gap sources using parameter $\gamma$:
\begin{equation}
\Delta_h = \gamma \Delta, \quad \Delta_d = \sqrt{1-\gamma^2} \Delta
\end{equation}
As indicated in the text we make the tight-binding simulation $4$-fold degenerate to mimic the valley/spin degeneracy present in an actual TLG system.

\section{TLG - details of the model}
For the description of the TLG bandstructure and eigenstates we employ the effective continuum Hamiltonian \cite{chenEvidenceGatetunableMott2019,zhangBandStructureABC2010, koshinoTrigonalWarpingBerry2009}, together with its associated notation and numerical values of simulation parameters. It consists of two parts $H_\mathrm{TLG}=H_\mathrm{ABC} + V_M(\mathbf{r})$, the first one describing the ABC-stacked trilayer graphene and the second one being due to moir{\'e} potential of hBN substrate. The trilayer graphene part is given by:

\begin{equation}
H_\mathrm{ABC} = \frac{\nu_0^3}{t_1^2} \begin{pmatrix}0 & k_+^3 \\ k_-^3 & 0 \end{pmatrix} + \left( \frac{2 \nu_0 \nu_3 k^2}{t_1} +t_2 \right)\sigma_x \notag + \left( \frac{2 \nu_0 \nu_4 k^2}{t_1} - \Delta' \right)\sigma_0 + \left( \frac{3 \nu_0^2 k^2}{t_1^2} +t_2 \right)\Delta'' \sigma_0 - \Delta \sigma_z
\end{equation}
where $k_\pm = \xi k_x \pm i k_y$, $\xi=\pm 1$ for $K$ and $K'$ valleys, $\nu_n = \sqrt{3}/2 a t_n$, $a=0.246 \mathrm{~nm}$ is the carbon-carbon lattice spacing and $t_0=2.62 \mathrm{~eV}$, $t_1=0.358 \mathrm{~eV}$, $t_2=-0.0083 \mathrm{~eV}$, $t_4=0.293 \mathrm{~eV}$, $t_5=0.144 \mathrm{~eV}$, $\Delta'=0.0122 \mathrm{~eV}$, $\Delta''=-0.0095 \mathrm{~eV}$, and $\kappa = 3.03$. Finally, the gaps in the bandstructure are opened and controlled by the applied electric field, which is described by $\Delta=50\, \mathrm{~meV}$ in our case to enlarge the range of energies between the intra- and interband continua. The moir{\'e} potential is given by:
\begin{equation}
V_M(\mathbf{r}) = 2 C_A \mathrm{Re}(e^{i \phi_A} f(\mathbf{r})) \begin{pmatrix} 1 & 0 \\ 0 & 0 \end{pmatrix}
\end{equation}
where $f(\mathbf{r}) = \sum_{j=1}^6 e^{i \mathbf{q}_j \mathbf{r}} (1 + (-1)^j)/2$ and $\mathbf{q}_j$ are the reciprocal lattice vectors of the triangular moir{\'e} supperlattice. The parameters used in the calculation of this potential are $C_A = -14.88 \mathrm{~meV}$ and $\phi_A=50.19\degree$. We obtain the energies and the eigenstates of $H_\mathrm{TLG}$ by numerical diagonalization using a momentum cutoff $q_C = 5 |\mathbf{q}_j|$. In all the calculations we are summing the results over both valleys and we take into account the 10 bands that lie the closest to the Fermi energy in order to consider all the relevant interband transitions. The Bloch wavefunction for a valley $\xi$ is taken as
\begin{equation}
\Psi_{\xi,n,\vec{k}}^{X}(\vec{r})=\sum_{\vec{G}} C_{\xi,n,\vec{k}}^{X}(\vec{G})e^{i(\vec{k}+\vec{G})\cdot \vec{r}}\label{eq:app_bloch_ansatz}
\end{equation}
with $X$ corresponding to each of the spinor components $X=A,B$. The band index is labeled by $n$ and $\vec{k}$ is the Bloch wave vector in the moir\'e superlattice Brillouin zone. Here $\vec{G}$ runs over all the possible integer combinations of the reciprocal lattice vectors, $\vec{G}=m_1 \vec{G}_{1}^{M}+ m_2 \vec{G}_{2}^{M}$ with integers $m_1$ and $m_2$ that satisfy the momentum cutoff condition.

 In order to evaluate the polarization function for a realistic bandstructure model \cite{zhangBandStructureABC2010, koshinoTrigonalWarpingBerry2009} it is necessary to slightly generalize the definition of the polarization function from Eq.\eqref{eq:app_pol_tight_binding}. The only required changes are: inclusion of multiple electron and hole bands in both $K$ and $K'$ valleys of TLG and a change in the definition of the band overlap factors $F_{\vec{k}+\vec{q},\vec{k}}^{nm}$:
\begin{equation}
F_{\vec{k}+\vec{q},\vec{k}}^{nm} = \left| \int_\Omega d^2r \Psi^\dagger_{n, \vec{k}+\vec{q}}(\vec{r}) e^{i \vec{q}\cdot\vec{r}} \Psi_{m,\vec{k}}(\vec{r})  \right|^2
\end{equation}
In the above expression, $n, m$ run over all the bands in both valleys, $\Psi_{m,\vec{k}}(\vec{r})$ represents the Bloch wave function for band $m$ and the integration is over moir\'e unit cell $\Omega$.

This model is spin-degenerate, which is taken into account by including a multiplying factor of 2 in the polarization function. With such changes to the polarization function we can numerically evaluate the dielectric function and determine plasmon dispersion from its nodes. 

\section{Electromagnetic field components of the plasmon waves}
\label{app:full_wave}

\begin{figure*}[!tb]
    \centering
    \includegraphics[width=\linewidth]{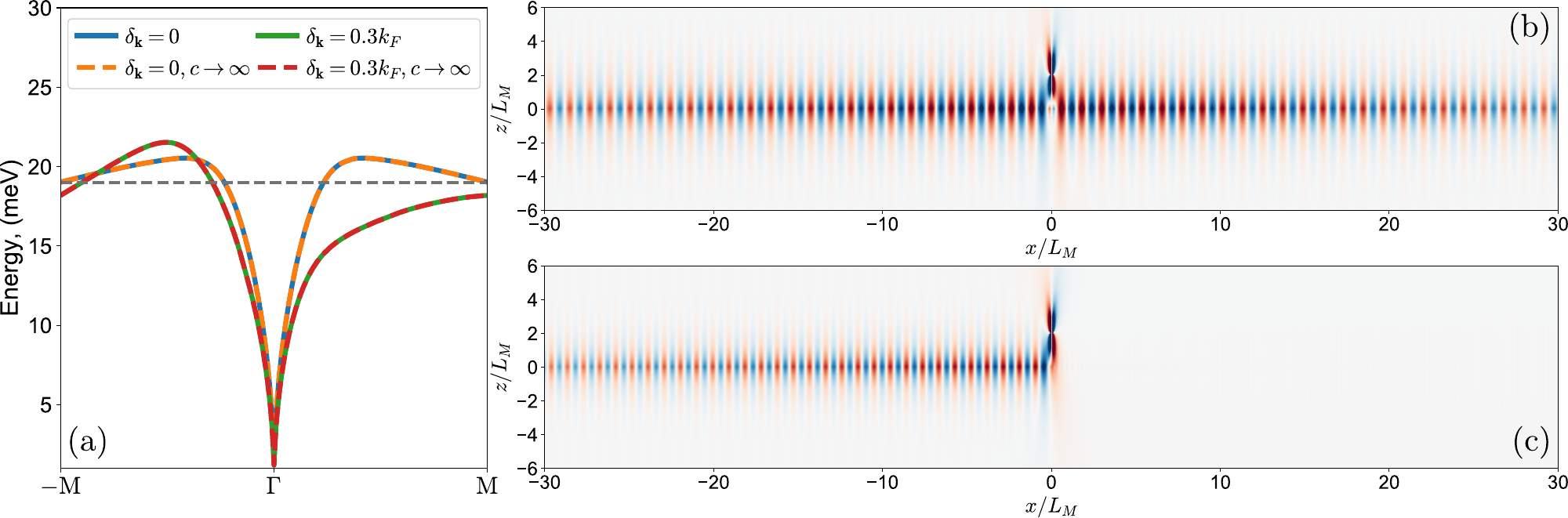}
    \caption{(a) Plasmon dispersion obtained as a solution of Eq.\eqref{eq:app_spp_dispersion} (solid line) and as nodes of Eq.\eqref{eq:app_dielectric_fun_def} (dashed line). The difference due to neglecting retardation effects is negligible. Horizontal dashed line indicates the energy of plasmons presented in panels (b) and (c). Time snapshot of electric field $E_x$ obtained using full-wave simulation (a) without and (b) with electric current.}
    \label{fig:fig_4}
\end{figure*}

In this Section of Supplemental Materials we discuss the behavior of the electromagnetic fields that comprise the surface plasmons for two different situations: a hybridized system without and with nonreciprocity in plasmons’ dispersion. We also demonstrate the validity of the non-retardation assumption for computing the dispersion of plasmons through seeking the nodes of the dielectric function from Eq. \eqref{eq:app_dielectric_fun_def}.

We seek to simulate the electric and magnetic fields that comprise the surface plasmons. In doing so we follow the analysis of Ref.\cite{doi:10.1063/1.2891452,acsphotonics.8b00987} and we determine electromagnetic response of an 2D electronic system to a current dipole located above it. We assume an infinite dipole line in the direction perpendicular to the current drift with the polarization in the direction perpendicular to the 2D sheet. In the simulation we place the dipole at a distance $2 L_M$ above the 2D sheet, where $L_M = 15$ nm is the moir{\'e} superlattice period. We take the dipole to emit radiation at a constant frequency corresponding to the plasmon frequency $\omega_p$. With this geometry in place we solve Maxwell Equations to obtain the behavior of both magnetic and electric fields in the whole space. 

In the movies 1, 2 and in the Fig. \ref{fig:fig_4} we plot the behavior of electric fields in a system for two different situations: a hybridized system without (b) and with (c) nonreciprocity in plasmons’ dispersion. For both situations we excite plasmon waves of the same frequency $\omega_p = 19$ meV, indicated by a gray line in Fig. \ref{fig:fig_4}(a). In the first case we find plasmons propagating away from the dipole in both positive and negative directions.  A snapshot of the propagation is shown in Fig. \ref{fig:fig_4}(b) with a full simulation included in the movie 1.  When both conventional and quantum Doppler effects are present Fig. \ref{fig:fig_4}(c), we find a plasmon mode which propagates in one direction only as clearly seen in movie 2. This unidirectionality is expected at plasmon frequencies where there is only one sign of the phase velocity. We stress that here this non-reciprocity arises, as we discussed in the main text, only because of the strong band hybridization and strong electron-electron interactions that are present in moir{\'e} materials. If only a conventional Doppler effect was present, at these drift velocities, we would find plasmons propagating away from the dipole in both positive and negative directions but with a small difference in the left and the right propagating wavelengths.  In the conventional Doppler effect, again at these realistic drift velocities, the range of plasmon frequencies that supports unidirectional propagation is extremely narrow.

We conclude this Section of the Supplemental Materials by demonstrating that there is no difference in the dispersion of the surface plasmons determined with and without the non-retardation assumption. The defining equation for the dispersion of transverse magnetic surface plasmon polariton (TM-SPP) waves is\cite{PhysRevB.80.245435}
\begin{equation}
\frac{\epsilon_{r1}}{\sqrt{q^2-\frac{\epsilon_{r1}\omega^2}{c^2}}}+\frac{\epsilon_{r2}}{\sqrt{q^2-\frac{\epsilon_{r2}\omega^2}{c^2}}}=-4\pi \kappa V_q \Pi(\omega,\vec{q})\label{eq:app_spp_dispersion}
\end{equation}
where $\epsilon_{r1}$, $\epsilon_{r2}$ are the dielectric constants of the materials encapsulating the 2D system, $\kappa=(\epsilon_{r1}+\epsilon_{r2})/2$ is their average, $V_q$ is the Coulomb potential as defined in the main text, $c$ corresponds to the speed of light and finally $\Pi(\omega,\vec{q})$ is the polarization function. If we assume that retardation effects can be neglected, i.e. $q \gg \omega/c$, then the above Eq.\eqref{eq:app_spp_dispersion} reduces to the form that stems from seeking the nodes of the Eq.\eqref{eq:app_dielectric_fun_def}. As mentioned in the main text, by seeking the nodes of the dielectric function, we therefore determine the dispersion of both conventionally defined plasmons (longitudinal charge oscillations) and the TM-SPP waves under condition that the retardation effects can be neglected. We verify this explicitly, as shown in Fig. \ref{fig:fig_4}(a), by solving Eq.\eqref{eq:app_spp_dispersion} for the dispersion of plasmons.

\end{document}